\documentclass[iop]{emulateapj}
\usepackage{subfigure}
\usepackage{multirow}
\usepackage[usenames,dvipsnames]{xcolor}
\usepackage[yyyymmdd,hhmmss]{datetime}

\makeatletter

\def\macs1149{MACS\,J1149.5+2223}

\def\obsdateone{30 October 2015}
\def\obsdatetwo{14 November 2015}
\def\obsdatethree{11 December 2015}

\def\sxcolor{0.2$\pm$0.3\,mag~AB}

\def\snrefsdal{SN\,Refsdal}

\def\coordsx{$\alpha$ = 11$^{\rm h}$49$^{\rm m}$36.02$^{\rm s}$, $\delta$ = +22$^{\circ}$23$'$48.1$''$ (J2000.0)}

\begin{document}
\title{D\'ej\`a Vu All Over Again: The Reappearance of Supernova Refsdal}

\shorttitle{Reappearance of Supernova Refsdal}

\newcommand{\HubbleFellow}{Hubble Fellow}
\newcommand{\ANU}{The Research School of Astronomy and Astrophysics, Australian National University, Mount Stromlo Observatory, via Cotter Road, Weston Creek, Australian Capital Territory 2611, Australia}
\newcommand{\Packard}{Packard Fellow}
\newcommand{\CalTech}{California Institute of Technology, 1200 East California Boulevard, Pasadena, CA 91125}
\newcommand{\Cantabria}{IFCA, Instituto de F\'isica de Cantabria (UC-CSIC), Av. de Los Castros s/n, 39005 Santander, Spain}
\newcommand{\IFCA}{\Cantabria}
\newcommand{\JHU}{Department of Physics and Astronomy, The Johns Hopkins University, 3400 N. Charles St., Baltimore, MD 21218, USA}
\newcommand{\Michigan}{Department of Astronomy, University of Michigan, 1085 S. University Avenue, Ann Arbor, MI 48109, USA}
\newcommand{\UCDavis}{University of California Davis, 1 Shields Avenue, Davis, CA 95616}
\newcommand{\UCLA}{Department of Physics and Astronomy, University of California, Los Angeles, CA 90095}
\newcommand{\USC}{Department of Physics and Astronomy, University of South Carolina, 712 Main St., Columbia, SC 29208, USA}
\newcommand{\RCEU}{Research Center for the Early Universe, University of Tokyo, 7-3-1 Hongo, Bunkyo-ku, Tokyo 113-0033, Japan}
\newcommand{\TokyoPhys}{Department of Physics, University of Tokyo, 7-3-1 Hongo, Bunkyo-ku, Tokyo 113-0033, Japan}
\newcommand{\IPMU}{Kavli Institute for the Physics and Mathematics of the Universe (Kavli IPMU, WPI), University of Tokyo, 5-1-5 Kashiwanoha, Kashiwa, Chiba 277-8583, Japan}
\newcommand{\TokyoAstro}{Department of Astronomy, Graduate School of Science, The University of Tokyo, 7-3-1 Hongo, Bunkyo-ku, Tokyo 113-0033, Japan}
\newcommand{\DARK}{Dark Cosmology Centre, Niels Bohr Institute, University of Copenhagen, Juliane Maries Vej 30, DK-2100 Copenhagen, Denmark} 
\newcommand{\INFN}{INFN, Sezione di Bologna, Viale Berti Pichat 6/2, I-40127 Bologna, Italy}
\newcommand{\EHU}{Fisika Teorikoa, Zientzia eta Teknologia Fakultatea, Euskal Herriko Unibertsitatea UPV/EHU}
\newcommand{\Basque}{IKERBASQUE, Basque Foundation for Science, Alameda Urquijo, 36-5 48008 Bilbao, Spain}
\newcommand{\Berkeley}{Department of Astronomy, University of California, Berkeley, CA 94720-3411, USA}
\newcommand{\STScI}{Space Telescope Science Institute, 3700 San Martin Dr., Baltimore, MD 21218, USA}
\newcommand{\Ferrara}{Dipartimento di Fisica e Scienze della Terra, Universit\`{a} degli Studi di Ferrara, via Saragat 1, I-44122, Ferrara, Italy}
\newcommand{\INAF}{INAF, Osservatorio Astronomico di Trieste , via G.B. Tiepolo 11, I-40131 Trieste, Italy}
\newcommand{\UCSB}{Department of Physics, University of California, Santa Barbara, CA 93106-9530, USA}
\newcommand{\SantaBarbara}{\UCSB}
\newcommand{\Kapteyn}{Kapteyn Astronomical Institute, University of Groningen, Postbus 800, 9700 AV Groningen, the Netherlands}
\newcommand{\WKU}{Department of Physics, Western Kentucky University, Bowling Green, KY 42101, USA}
\newcommand{\IAP}{Institut d’Astrophysique de Paris, UMR7095 CNRS-Universit\'{e} Pierre et Marie Curie, 98bis bd Arago, F-75014 Paris, France}
\newcommand{\ASIAA}{Institute of Astronomy and Astrophysics, Academia Sinica, P.O. Box 23-141, Taipei 10617, Taiwan}
\newcommand{\TokyoKashiwa}{Institute for Cosmic Ray Research, The University of Tokyo, Kashiwa, Chiba 277-8582, Japan}
\newcommand{\Munich}{University Observatory Munich, Scheinerstrasse 1, D-81679 Munich, Germany} 
\newcommand{\KICPStanford}{Kavli Institute for Particle Astrophysics and Cosmology, Stanford University, 452 Lomita Mall, Stanford, CA 94305, USA}
\newcommand{\Andalucia}{Instituto de Astrof\'isica de Andaluc\'ia (CSIC), E-18080 Granada, Spain}
\newcommand{\SaoPaulo}{Instituto de Astronomia, Geof\'isica e Ci\^encias Atmosf\'ericas, Universidade de S\~ao Paulo, Cidade Universit\'aria, 05508-090, S\~ao Paulo, Brazil}
\newcommand{\AMNH}{Department of Astrophysics, American Museum of Natural History, Central Park West and 79th Street, New York, NY 10024, USA}
\newcommand{\NYU}{Center for Cosmology and Particle Physics, New York University, New York, NY 10003, USA}
\newcommand{\Arizona}{Department of Astronomy, University of Arizona, Tucson, AZ 85721, USA}
\newcommand{\Rutgers}{Department of Physics and Astronomy, Rutgers, The State University of New Jersey, Piscataway, NJ 08854, USA}
\newcommand{\NOAO}{National Optical Astronomical Observatory, Tucson, AZ 85719, USA}
\newcommand{\LCOGT}{Las Cumbres Observatory Global Telescope Network, 6740 Cortona Dr., Suite 102, Goleta, California 93117, USA}
\newcommand{\IllinoisAstro}{ Astronomy Department, University of Illinois at Urbana-Champaign, 1002 W.\ Green Street, Urbana, IL 61801, USA }
\newcommand{\IllinoisPhysics}{ Department of Physics, University of Illinois at Urbana-Champaign, 1110 W.\ Green Street, Urbana, IL 61801, USA }
\newcommand{\GSFC}{Astrophysics Science Division, NASA Goddard Space Flight Center, MC 661, Greenbelt, MD 20771, USA}
\newcommand{\UMD}{Joint Space-Science Institute, University of Maryland, College Park, MD 20742, USA}

\newcommand{\StonyBrook}{Physics and Astronomy Department, Stony Brook University, Stony Brook, NY 11794-3800 }
\newcommand{\AIP}{Leibniz-Institut f\"ur Astrophysik Potsdam (AIP), An der Sternwarte 16, 14482 Potsdam, Germany}

\newcounter{affilct}
\setcounter{affilct}{0}

\makeatletter
\newcommand{\affilref}[1]{%
  \@ifundefined{c@#1}%
    {\newcounter{#1}%
     \setcounter{#1}{\theaffilct}%
     \refstepcounter{affilct}%
     \label{#1}%
     }{}%
  \ref{#1}%
 }
\makeatother

\makeatletter
\newcommand*\affilreftxt[2]{%
  \@ifundefined{c@#1txt}
    {\newcounter{#1txt}%
     \setcounter{#1txt}{1}
     \altaffiltext{\ref{#1}}{#2}
     }{
     }
  }
\makeatother

\author{P.~L.~Kelly\altaffilmark{\affilref{Berkeley}}}
\affilreftxt{Berkeley}{\Berkeley}
\email{pkelly@astro.berkeley.edu}

\author{S.~A.~Rodney\altaffilmark{\affilref{USC},}}
\affilreftxt{USC}{\USC}

\author{T.~Treu\altaffilmark{\affilref{UCLA},\affilref{Packard}}}
\affilreftxt{UCLA}{\UCLA}
\affilreftxt{Packard}{\Packard}

\author{L.-G.~Strolger\altaffilmark{\affilref{STScI}}}
\affilreftxt{STScI}{\STScI}

\author{R.~J.~Foley\altaffilmark{\affilref{IllinoisPhysics},\affilref{IllinoisAstro}}}
\affilreftxt{IllinoisPhysics}{\IllinoisPhysics}
\affilreftxt{IllinoisAstro}{\IllinoisAstro}

\author{S.~W.~Jha\altaffilmark{\affilref{Rutgers}}}
\affilreftxt{Rutgers}{\Rutgers}

\author{J.~Selsing\altaffilmark{\affilref{DARK}}}
\affilreftxt{DARK}{\DARK}

\author{G.~Brammer\altaffilmark{\affilref{STScI}}}
\affilreftxt{STScI}{\STScI}

\author{M.~Brada\v{c}\altaffilmark{\affilref{UCDavis}}}
\affilreftxt{UCDavis}{\UCDavis}

\author{S.~B. Cenko\altaffilmark{\affilref{GSFC},\affilref{UMD}}}
\affilreftxt{GSFC}{\GSFC}
\affilreftxt{UMD}{\UMD}

\author{O.~Graur\altaffilmark{\affilref{NYU},\affilref{AMNH}}}
\affilreftxt{NYU}{\NYU}
\affilreftxt{AMNH}{\AMNH}

\author{A.~V.~Filippenko\altaffilmark{\affilref{Berkeley}}}
\affilreftxt{Berkeley}{\Berkeley}

\author{J.~Hjorth\altaffilmark{\affilref{DARK}}}
\affilreftxt{DARK}{\DARK}

\author{C.~McCully\altaffilmark{\affilref{LCOGT},\affilref{UCSB}}}
\affilreftxt{LCOGT}{\LCOGT}
\affilreftxt{UCSB}{\UCSB}

\author{A.~Molino\altaffilmark{\affilref{SaoPaulo},\affilref{Andalucia}}}
\affilreftxt{Andalucia}{\Andalucia}
\affilreftxt{SaoPaulo}{\SaoPaulo}

\author{M.~Nonino\altaffilmark{\affilref{INAF}}}
\affilreftxt{INAF}{\INAF}

\author{A.~G.~Riess\altaffilmark{\affilref{JHU},\affilref{STScI}}}
\affilreftxt{JHU}{\JHU}
\affilreftxt{STScI}{\STScI}

\author{K.~B.~Schmidt\altaffilmark{\affilref{UCSB},\affilref{AIP}}}
\affilreftxt{UCSB}{\UCSB}
\affilreftxt{AIP}{\AIP}

\author{B.~Tucker\altaffilmark{\affilref{ANU}}}
\affilreftxt{ANU}{\ANU}

\author{A.~von~der~Linden\altaffilmark{\affilref{StonyBrook}}}
\affilreftxt{StonyBrook}{\StonyBrook}

\author{B.~J.~Weiner\altaffilmark{\affilref{Arizona}}}
\affilreftxt{Arizona}{\Arizona}

\author{A.~Zitrin\altaffilmark{\affilref{CalTech},\affilref{HubbleFellow}}}
\affilreftxt{CalTech}{\CalTech}

\keywords{gravitational lensing: strong --- supernovae: general, individual: SN
Refsdal --- galaxies --- clusters: general, individual: MACS\,J1149.5+2223}

\begin{abstract}
In {\it Hubble Space Telescope (HST)} imaging taken on 10 November 2014, four images of supernova (SN) ``Refsdal'' (redshift $z = 1.49$) appeared in an Einstein-cross--like configuration (images S1--S4) around an early-type galaxy in the cluster MACS J1149.5+2223 ($z = 0.54$). 
Almost all lens models of the cluster have predicted that the SN should reappear within a year in a second host-galaxy image created by the cluster's potential. In {\it HST} observations taken on 11 December 2015, we find a new source at the predicted position of the new image of SN Refsdal approximately $8''$ from the previous images S1--S4. This marks the first time the appearance of a SN at a particular time and location in the sky was successfully predicted in advance! We use these data and the light curve from the first four observed images of SN Refsdal to place constraints on the relative time delay and magnification of the new image (SX), compared to images S1--S4. This enables us, for the first time, to test ``blind'' lens model predictions of both magnifications and time delays for a lensed SN. We find that the timing and brightness of the new image are consistent with the blind predictions of a fraction of the models. The reappearance illustrates the discriminatory power of this blind test and its utility to uncover sources of systematic uncertainty. From planned {\it HST} photometry, we expect to reach a precision of 1--2\% on the time delay between S1--S4 and SX.
\end{abstract}


\maketitle 

\section{Introduction}
Background sources strongly lensed by galaxies and galaxy clusters
that exhibit flux variations in time can be used as powerful probes,
because they make it possible to
measure the relative time delays between their multiple images. 
As \citet{refsdal64} first suggested, time delays are useful 
because they depend sensitively on both the cosmic expansion rate 
and the gravitational potential of the lens. 
While the positions of the images of lensed galaxies depend on the derivative of
the potential, time delays are directly proportional to differences in the
potential.

\citet{refsdal64} examined the utility of time-delay measurements of
a strongly lensed, multiply imaged, resolved supernova (SN), but such an object
was not found in the following five decades.  A handful of SN in galaxy-cluster fields have been magnified ($\sim1.4$--4$\times$) by the cluster's potential \citep{Goobar:2009,Patel:2014,nordinrubinrichard14,rodneypatelscolnic15}, 
but none has been multiply imaged.
A luminous H-poor SN at redshift $z=1.38$ \citep{chornockbergerrest13} was shown to be a highly magnified ($\sim30\times$) SN~Ia \citep{quimbywerneroguri13,quimbyoguirmore14}, but the only existing exposures, taken from the ground, could not resolve multiple images of the SN.

Although strongly lensed SN have eluded detection for 50 years, the
discovery of multiply imaged quasars beginning in 1979 \citep{Wal++79} has 
made it possible to measure time delays for more
than 20 systems \citep[see, e.g.,][]{Kun++97,Fas++99,Tew++13}.
For a subset of multiply imaged quasars with simple, early-type galaxy
lenses, it has been possible to precisely predict the delay arising
from the gravitational potential and thereby to measure an absolute distance 
scale and $H_0$ geometrically \cite[e.g.,][]{Paraficz:2010,Suy++13,Suy++14}.  Although
they are more difficult to find, strongly lensed SN hold great promise
as tools for time-delay cosmography
\citep{Kolatt:1998,Holz:2001,Bolton:2003,ogurikawano03,doblerkeeton06}.  In comparison with
quasars light curves, those of SN are relatively simple, and the peak
luminosities of SN~Ia \citep{ph93} and (with less precision) SN~IIP \citep[e.g.,][]{kirshnerkwan74} can be calibrated absolutely, thus providing a measurement
of lensing magnification.

\citet{kellyrodneytreu15} reported the discovery of \snrefsdal, the first
strongly lensed SN resolved into multiple images, in the \macs1149
\citep{ebelingedgehenry01,ebelingbarrettdonovan07} galaxy cluster
field in {\it Hubble Space Telescope (HST)} images taken on 10 November 2014 (UT dates are used throughout this paper).
Those exposures, collected as part of the Grism Lens-Amplified Survey from Space
(GLASS; PI Treu; GO-13459; \citealt{schmidttreubrammer14};
\citealt{treuschmidtbrammer15}), revealed four resolved
images of the background SN arranged in an Einstein cross
configuration around an elliptical cluster member.  Models of the
complex potential of the galaxy cluster and early-type galaxy lens
suggest that three of the four images are magnified by up to a factor
of $\sim10$--20 \citep{kellyrodneytreu15,oguri15,sharonjohnson15,
diegobroadhurstchen16,grillokarmansuyu15,jauzacrichardlimousin15,
kawamataoguriishigaki15,treubrammerdiego16}.

Models of the massive \macs1149\ cluster [($1.4 \pm 0.3) \times 10^{15}$ M$_{\odot}$; \citealt{vdlallen14,kellyvonderlinden14,applegatevdl14}]  produced soon after the discovery predicted that 
\snrefsdal\ would appear within several years in 
a different host-galaxy image close to the cluster core, $\sim8''$ from images S1--S4 \citep{kellyrodneytreu15,oguri15,sharonjohnson15,diegobroadhurstchen16}.
We adopt the identifier ``SX'' for this new image, following
\citet{oguri15}.  Here we report the appearance of the image SX of SN Refsdal in {\it HST} images (PI: Kelly; GO-14199) taken on \obsdatethree.

In models of the \macs1149\ cluster lens, the galaxy-cluster gravitational
potential is constrained by varying combinations of strong-lensing
constraints, including the positions and redshifts of multiply imaged
background galaxies, the positions of the \snrefsdal\ images S1--S4,
and locations of bright clumps within \snrefsdal's host galaxy. The
potential (or surface mass density) is also parameterized in a variety
of ways, often using the positions or light distributions of cluster
galaxies as constraints.
Given the complexity of the cluster potential, it is unlikely that
measurement of time delays between the \snrefsdal\ images can be used
for precision cosmology as suggested by \citet{refsdal64}. However,
if one adopts a fixed set of cosmological parameters, then time delays
and magnification ratios can be used to measure the difference in the
potential and its derivatives between the positions of multiple
images, thus providing a powerful local test of lens models.

To sharpen this test, several lens modeling teams have refined their
predictions for the relative time delay and magnification of image SX.
\citet{treubrammerdiego16} identified an improved set of multiply
imaged galaxies using additional data collected soon before and after the discovery of
the SN.  Systems were discovered or confirmed from {\it HST} WFC3 G102
and G141 grism spectra (PI Treu; GO-13459), thirty orbits of G141 grism spectra
taken to determine the spectroscopic type of the SN (PI Kelly;
GO-14041; \citealt{kellybrammerselsing15}; Brammer et al., in prep.), deep
VLT-MUSE observations (PI Grillo; Grillo et al. 2016, submitted),
Keck/DEIMOS observations (PI Jha), as well as Frontier Fields
observations of the \macs1149 field that began shortly after discovery
(PI Lotz; GO-13504). These data provided 429
spectroscopic redshifts in the field of MACSJ1149.5+2223, including
170 cluster members and 23 multiple images of 10
different galaxies. With the improved dataset,
\citet{treubrammerdiego16} organized 5 independent lens-modeling teams
which produced 7 separate predictions for the time delay.  In a
parallel effort, \citet{jauzacrichardlimousin15} used new Gemini GMOS
and part of the VLT-MUSE data (PI Grillo), as well as Frontier Fields
photometry, to generate improved constraints on the cluser potential
and new predictions for the time delay and magnification of image SX.

These revised models largely favored delays of less than one year
and predicted that image SX would be 
significantly fainter than images S1--S3, by a factor of 3--4.
Together, these predictions indicated that image SX could plausibly
have been detected as soon as {\it HST} could observe the \macs1149 field 
beginning on 30 October 2015. From late July through late October, it had been
too close to the Sun to be observed.  Importantly, all of these
modeling efforts were completed before the first realistic opportunity
to detect image SX on 30 October 2015, making these truly {\it blind}
predictions. 

Here we present a direct test of the lens model predictions, as we revisit the \macs1149 field and identify the appearance of the anticipated fifth image of \snrefsdal.  
In \citet{kellybrammerselsing15}, we classify SN Refsdal as the explosion of an H-rich compact massive star broadly similar to SN 1987A, and in \citet{rodneystrolgerkelly15} we measure time delays between images S1 through S4.
Section~\ref{sec:methods} in this paper presents the data processing
and photometry of the new {\it HST} images.  In
Section~\ref{sec:results}, we derive joint constraints on the relative
time delay and magnification of image SX and compare these to the published
predictions from the lens-modeling community. We briefly discuss our
results in Section~\ref{sec:discussion} and conclude in
Section~\ref{sec:summary}.  Throughout this paper, magnitudes are
given in the AB system \citep{okegunn83}, and a concordance cosmology
is assumed when necessary ($\Omega_m=0.3$, $\Omega_\Lambda=0.7$,
$H_0$ = 70 km s$^{-1}$ Mpc$^{-1}$).

\section{Methods}
\label{sec:methods}
We processed the WFC3 imaging data using a pipeline constructed from
the {\tt DrizzlePac} software
tools.\footnote{\url{http://drizzlepac.stsci.edu}} The images were
resampled to a scale of 0.06\arcsec~pix$^{-1}$ using {\tt
  AstroDrizzle} \citep{fruchter10} and then registered to a common
astrometric frame using {\tt TweakReg}.  Template images in each band were 
constructed by combining all available WFC3 infrared (IR) imaging collected
prior to 30 October 2015, comprising observations from the GLASS program,
the Cluster Lensing And Supernova survey with Hubble (CLASH; GO-12068;
PI M. Postman; \citealt{Postman:2012}), the Hubble Frontier Fields
(HFF; DD/GO-13504; PI J. Lotz), the FrontierSN program (GO-13790; PI
S. Rodney), and the \snrefsdal\ Follow-up program (DD/GO-14041; PI
P. Kelly).  The excellent stability of the {\it HST} point-spread function (PSF) allowed us to
generate difference images by simply subtracting these
archival template images directly from the search-epoch images.

To measure the SN flux from the difference images, we used the {\tt
  PythonPhot}\footnote{\url{https://github.com/djones1040/PythonPhot}}
software package \citep{Jones:2015} which implements PSF fitting
based on the {\tt DAOPHOT} algorithm
\citep{Stetson:1987}. We measure photometric uncertainties by
planting and recovering 1000 fake stars (copies of a model PSF) in
the vicinity of the SN position.  

A principal inference we make is to constrain the time delay and magnification of
image SX relative to image S1.
We simulate potential light curves of image SX by shifting the light curve of image S1 in time and demagnifying it.
To construct a simple model of the light curve of image S1, we fit separate second-order polynomials to its {\it F125W} and {\it F160W} flux measurements (see \citealt{rodneystrolgerkelly15} and \citealt{kellybrammerselsing15}).
Across a grid of time delays and magnifications, we next calculate the
expected brightness in {\it F125W} and {\it F160W} of image SX at each epoch listed in  Table~\ref{tab:photometry}.
The combined probability of the measured fluxes is taken to be the likelihood of each pair of relative time delay and magnification values.
While we use the light curve of image S1 as a model, those of S2 and S3 yield almost identical constraints.

\section{Results}
\label{sec:results} 
\begin{figure*}
\centering
\subfigure{ \includegraphics[angle=0,width=6.5in]{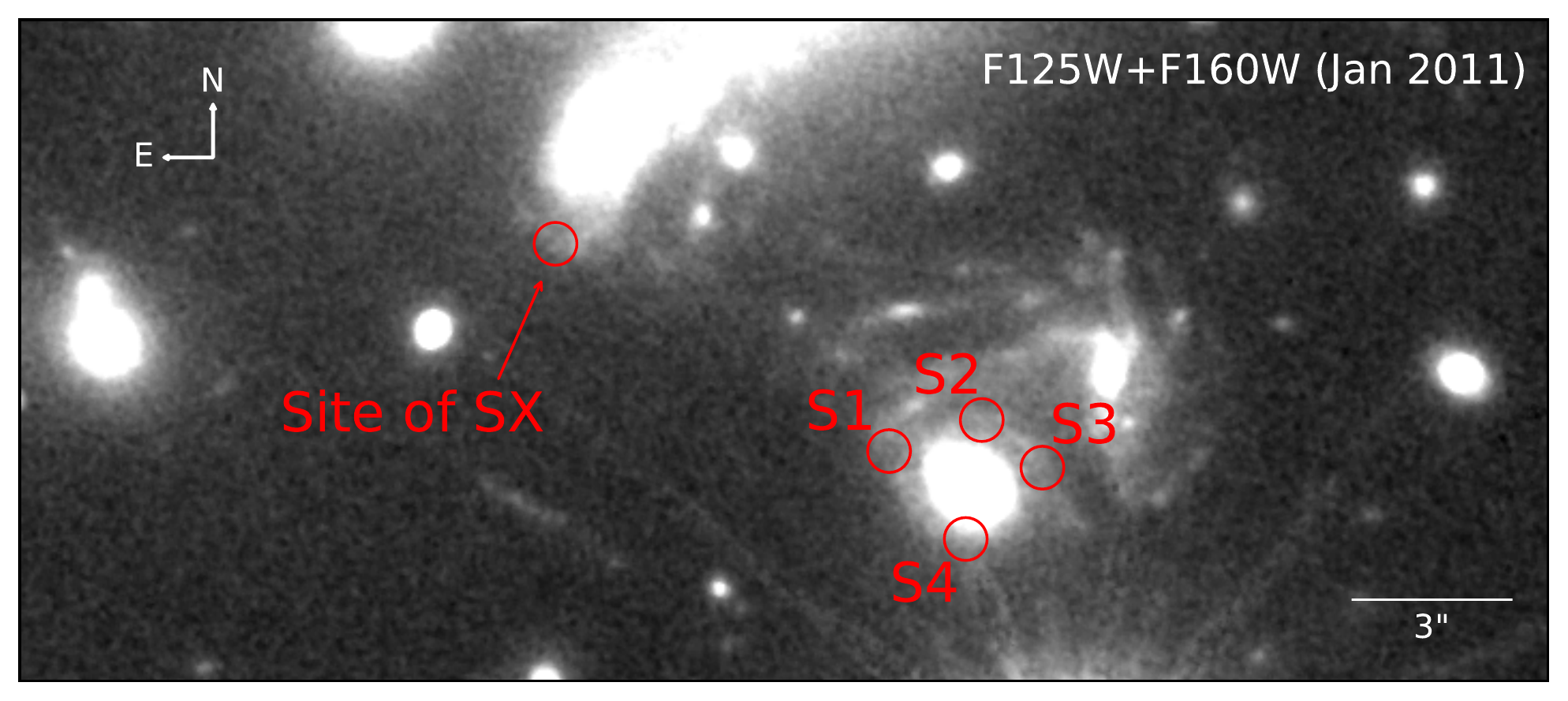} }
\subfigure{ \includegraphics[angle=0,width=6.5in]{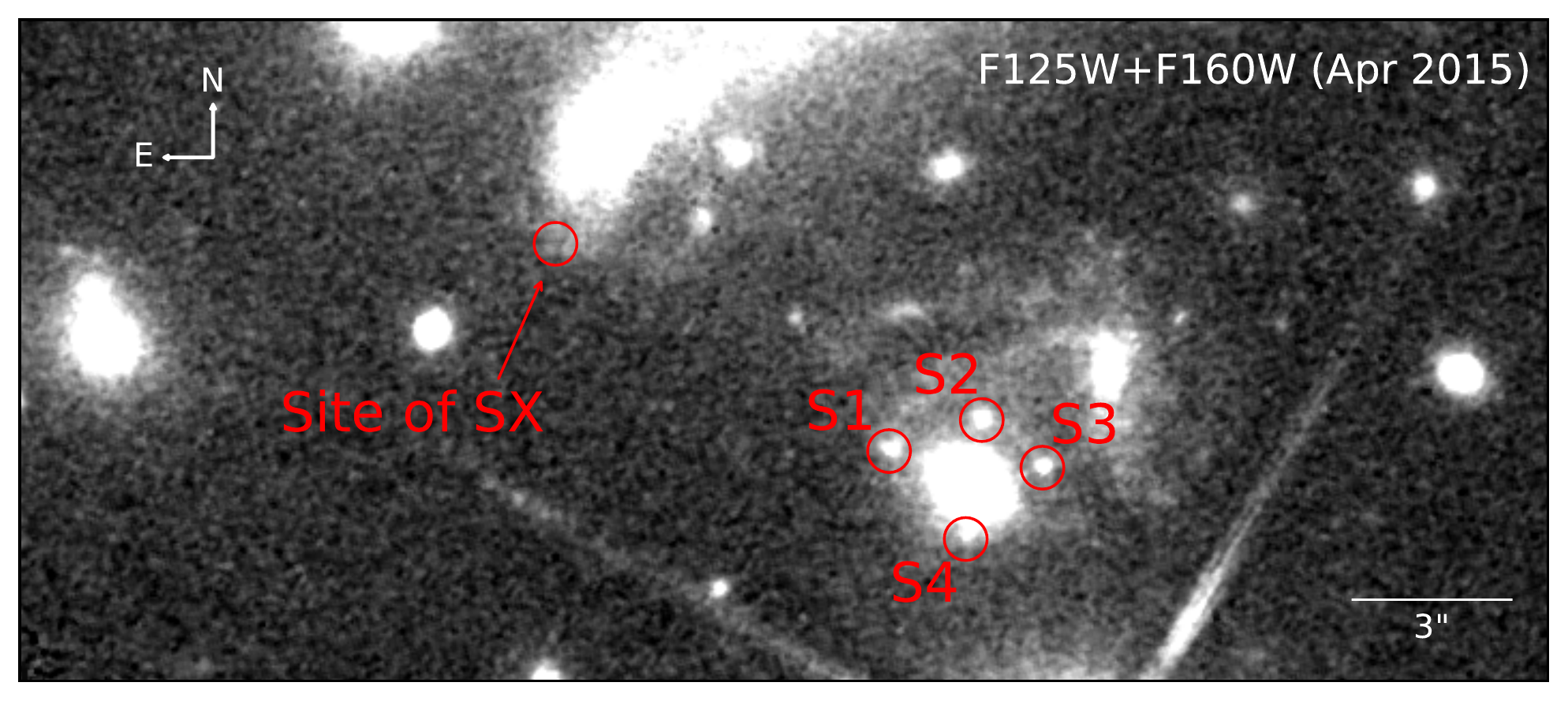} }
\subfigure{ \includegraphics[angle=0,width=6.5in]{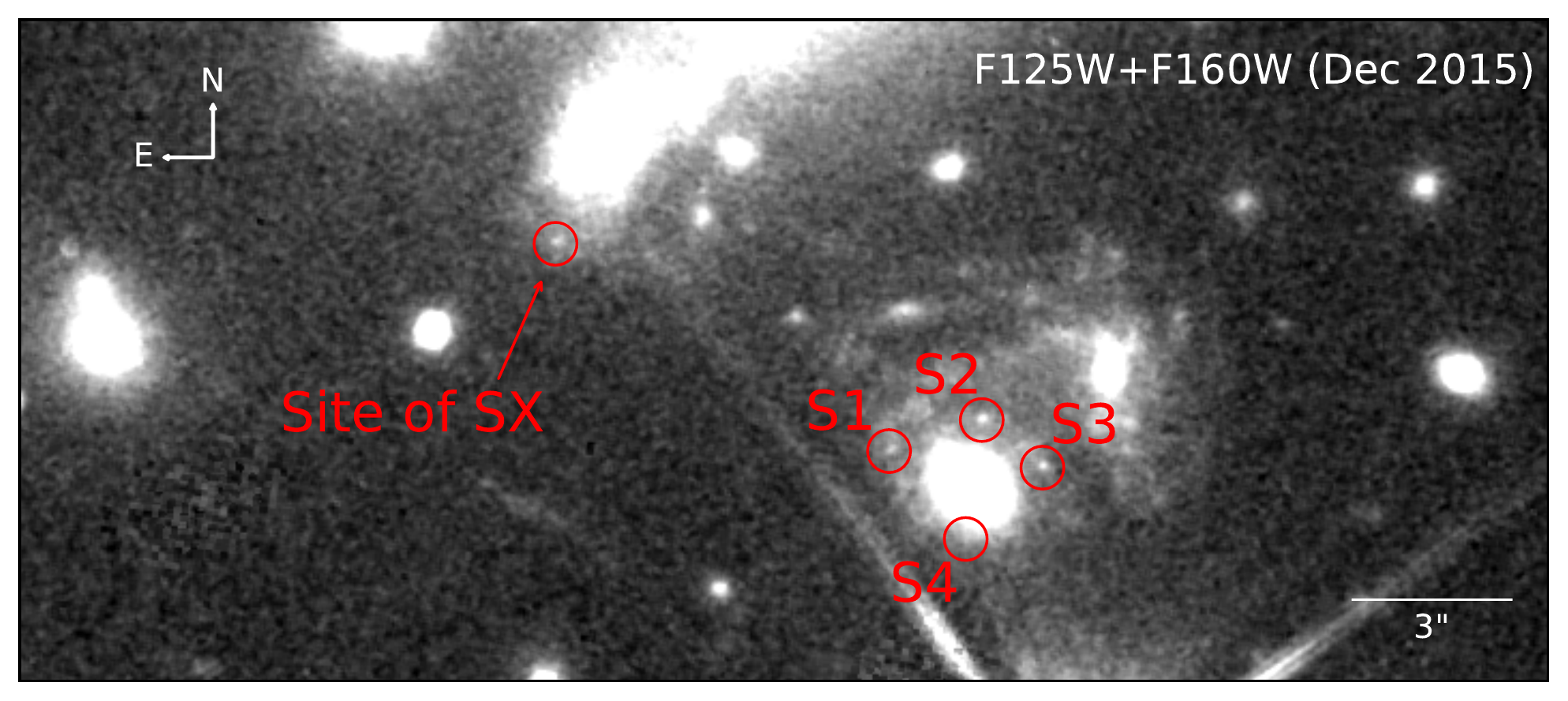} }
\caption{Coadded WFC3-IR {\it F125W} and {\it F160W} exposures of the
\macs1149 galaxy-cluster field taken with {\it HST}.
The top panel shows images acquired in 2011 before the SN appeared in S1--S4 or SX.
The middle panel displays images taken on 20 April 2015 when
the four images forming the Einstein cross are close to maximum brightness, but 
no flux is evident at the position of SX. 
The bottom panel shows images taken on \obsdatethree\ which reveal the new image SX of SN Refsdal.
Images S1--S3 in the Einstein cross configuration remain visible in the \obsdatethree\ coadded image (see \citealt{kellybrammerselsing15} and \citealt{rodneystrolgerkelly15} for analysis of the SN light curve).
}
\label{fig:imagereappear}
\end{figure*}

In Figure~\ref{fig:imagereappear}, we present a coaddition of the {\it F125W} and {\it
  F160W} images taken on \obsdatethree~which shows the reappearance of
\snrefsdal\ in a different image of its $z=1.49$ host galaxy.  
The coordinates of this image SX are
\coordsx.\footnote{The coordinates are registered to
  the astrometric system used for the CLASH, GLASS, and HFF images and
  catalogs, \url{http://www.stsci.edu/hst/campaigns/frontier-fields/} .}
This locates SX at 6.2\arcsec\ North and 3.9\arcsec\ East of image S1.
Table~\ref{tab:photometry} reports the measured fluxes and
uncertainties.  
We find a {\it F125W}\,$-$\,{\it F160W} color of \sxcolor~for image
SX, consistent with those measured for four SN images forming the Einstein cross at discovery.
We detect a fainter source at the same coordinates in a combination of the {\it F125W} and {\it F160W} images taken on \obsdatetwo, while the \obsdateone\ images yield no statistically significant detection (see Table~\ref{tab:photometry}).
Images S1--S3 of SN Refsdal remain visible in the coaddition of images taken on \obsdatethree.

\begin{deluxetable*}{ccccc}
  \tablecolumns{5}
  \tablecaption{\sc Measurements of Image SX \label{tab:photometry}}
  \tablecomments{The fluxes are measured at the position of image SX. 
  Photometry of the images obtained on \obsdatethree~yields a {\it F125W}\,$-$\,{\it F160W} color of \sxcolor. The coordinates of image SX are \coordsx. }
  \tablehead{
    \multicolumn{2}{c}{Obs. Date} & \colhead{Filter} & \colhead{Exp. Time} & \colhead{Magnitude} \\
    \colhead{} & \colhead{(MJD)} & \colhead{} & \colhead{(s)} & \colhead{(AB)} }
  \startdata
\obsdateone & 57325.8 & F125W &  1259 & 27.4 $\pm$ 0.4 \\
\obsdatetwo & 57340.9 & F125W &  1259 & 27.3 $\pm$ 0.4 \\
\obsdatethree & 57367.1 & F125W &  1259 & 26.56 $\pm$ 0.16 \\[1mm]
\obsdateone & 57325.9 & F160W &  1159 & 27.4 $\pm$ 0.6 \\
\obsdatetwo & 57341.0 & F160W &  1159 & 26.29 $\pm$ 0.15 \\
\obsdatethree & 57367.1 & F160W &  1159 & 26.24 $\pm$ 0.16 
\enddata
\end{deluxetable*}

\begin{figure*}
\centering
\subfigure{ \includegraphics[angle=0,width=5.5in]{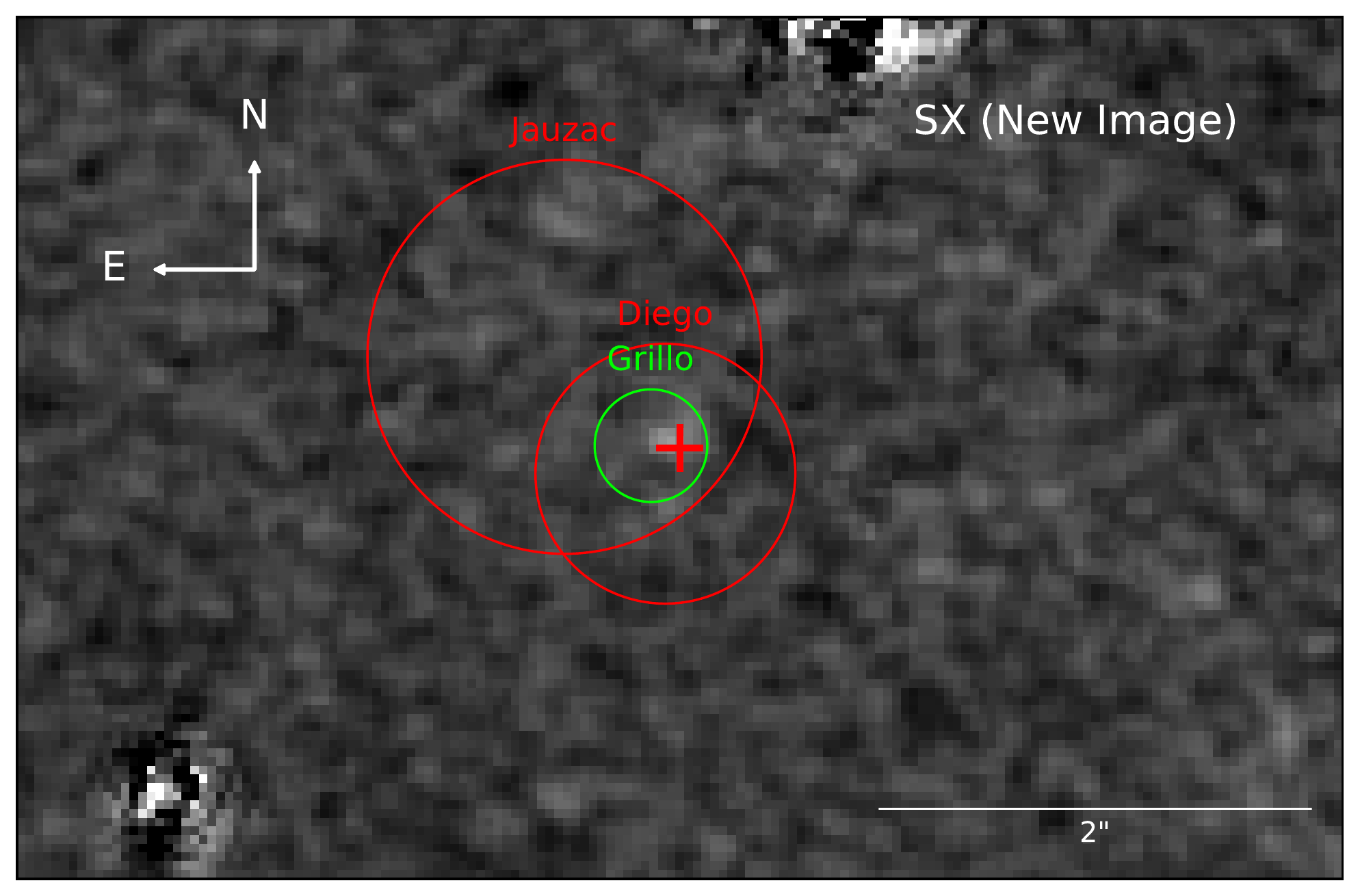} }
\caption{Comparison between the predicted and the actual position of image SX of SN Refsdal. 
Coordinate published predictions are overplotted on the coaddition of {\it F125W} and {\it F160W} difference images made by subtracting the 
\obsdatethree\ exposures from archival template images taken in 2011.
The circles show the root-mean square (rms) of the angular offsets between the measured positions of multiply imaged sources and their positions in the best-fitting respective models.  The \citet{diegobroadhurstchen16}, \citet{jauzacrichardlimousin15}, and \citet{grillokarmansuyu15} predictions are all consistent with the measured position of image SX within the reported rms scatter. The residual scatter for the \citet{diegobroadhurstchen16} model was not published and is 0.6$''$ (priv. comm.).
}
\label{fig:coordinates}
\end{figure*}

\begin{figure*}
\centering
\subfigure{ \includegraphics[angle=0,width=6.5in]{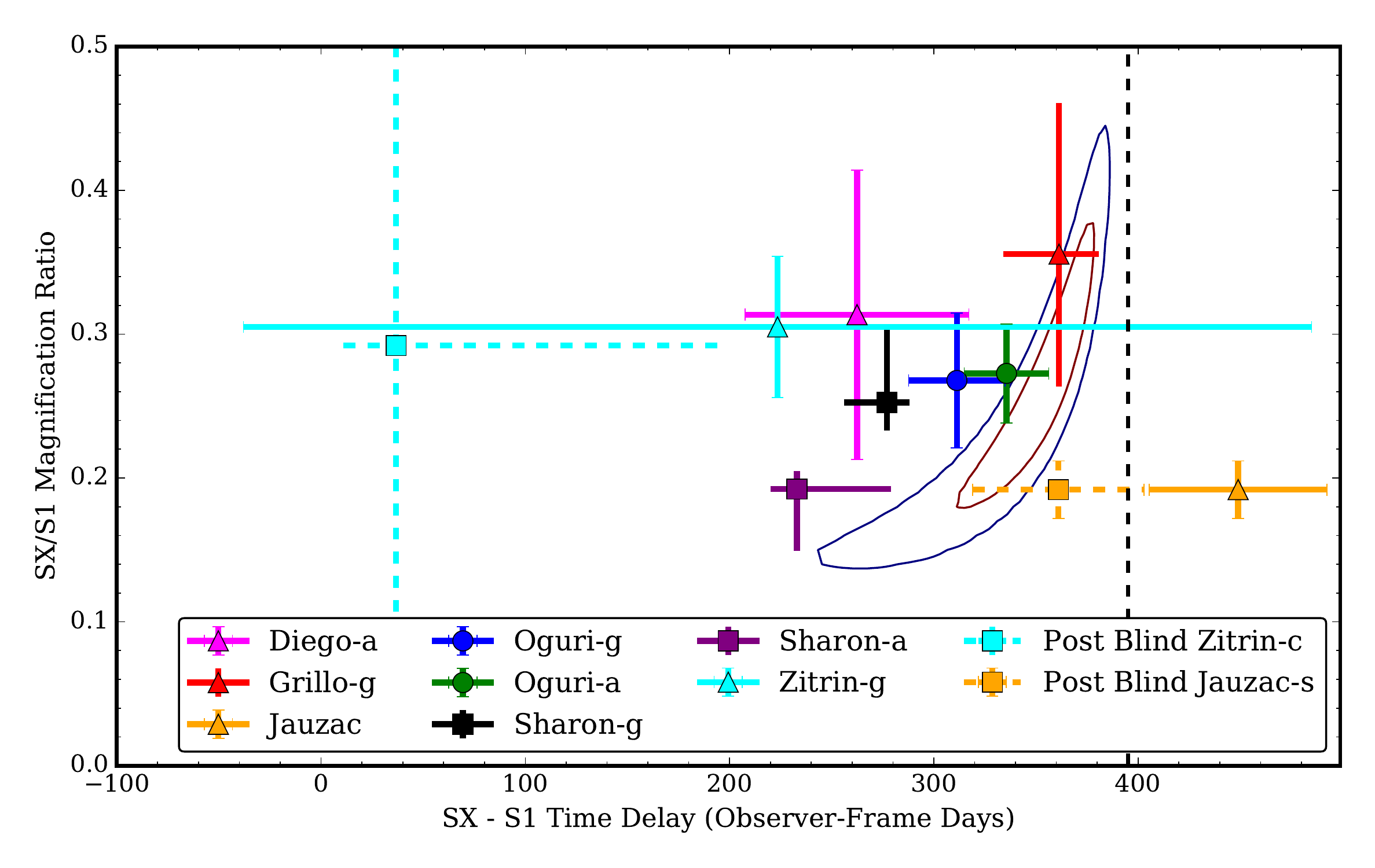} }
\caption{Simultaneous constraints on the time delay and magnification of image SX relative to image S1 from photometry of image SX listed in Table~\ref{tab:photometry}. 
The two-dimensional contours show the 
  68\% and 95\% confidence levels, and model predictions plot 68\% confidence levels. 
   Since many of the lensing predictions are not Gaussian distributed, the 68\%
  limits do not imply that they are necessarily inconsistent with the measurements. 
   Except for the \citet{jauzacrichardlimousin15} prediction, labels refer to models 
  presented by \citet{treubrammerdiego16}. 
  While all other plotted predictions were made in advance of the {\it HST} Cycle 23 observations beginning on \obsdateone, ``Post Blind Zitrin-c'' and ``Post Blind Jauzac'' were updates made at a later date. ``Post Blind Zitrin-c'' is an update of the ``Zitrin-g'' model where the lens galaxy was left to be freely weighted to assure that its critical curves pass between the four Einstein-cross images.
For ``Post Blind Jauzac,'' the authors compute a common position for images S1--S4 in the source plane and recompute the time delays analytically using their {\tt LENSTOOL} model of the cluster potential. 
  The greater the S1--SX delay, the earlier the \obsdatethree\ observations are in the light
curve of SX. The black dashed line marks the delay beyond which we lack data on the light curve of SN Refsdal. We extrapolate to earlier epochs using the best-fitting second-order polynomials.
  }
\label{fig:delaymu}
\end{figure*}

Figure~\ref{fig:coordinates} shows a comparison between the coordinates of the 
new image SX and several published model predictions, which are in good agreement. 
In Figure~\ref{fig:delaymu}, we plot simultaneous constraints on the time
delay and magnification ratio between image S1 discovered in 2014 and
the newly detected image SX.
We show model predictions for the relative time delay and magnification between SX and S1 from several teams reported by \citet{treubrammerdiego16} as well as
independent predictions by \citet{jauzacrichardlimousin15}.  

While the other plotted predictions were made blind to the observations beginning on \obsdateone, the two predictions plotted in Figure~\ref{fig:delaymu} with a dashed line, ``Zitrin-c'' and ``Jauzac-s,'' were updated at a later date;
hence, these last two are not truly blind predictions, although they incorporate no additional data.
``Zitrin-c,'' an improvement over the ``Zitrin-g'' model, allows the early-type lens galaxy to be freely weighted to assure that the critical curves pass between the four Einstein-cross images.
In the case of ``Jauzac-s,'' the authors addressed an issue with the use of {\tt LENSTOOL} \citep{kneibellissmail96,jullokneiblimousin07} for computing time delays. 
{\tt LENSTOOL} allows the multiple images of a strongly lensed object such as SN Refsdal to map to differing positions on the source plane, which can translate into an aberration of the predicted relative time delays.  For ``Jauzac-s,'' the authors estimate a single, common position for SN Refsdal in the source plane and analytically estimate improved model time delays using this common position (see Section 5.2.2 of \citealt{jauzacrichardlimousin15}).

\section{Discussion}
\label{sec:discussion}

Lensed SN provide a powerful means to test the accuracy of the lens
models of the foreground deflector, or to provide additional input
constraints \citep[e.g.,][]{Riehm:2011}. Previous tests have been based
on SN that are magnified but not multiply imaged
\citep{patelmccullyjha14,nordinrubinrichard14}. Recently,
\citet{rodneypatelscolnic15} discovered a Type Ia SN magnified by a
factor of $\sim2$ by a galaxy-cluster potential and found
that its calibrated luminosity was in tension with some --- but not all
--- models of the cluster potential.

With \snrefsdal\ we have for the first time been able to test
predictions for both the lensing time delay and the
magnification. This is important because the time delay depends on the
difference in gravitational potential, while magnification depends on
a combination of second derivatives, and therefore the two observables
test different aspects of the potential. In principle, time delays are
much less sensitive than magnification ratios to millilensing and
microlensing; they should therefore be more robustly predicted.

It is important to keep in mind that all of these tests are local, and thus a
larger sample is needed to assess the global goodness of fit of every
model. Nevertheless, these tests are an extremely valuable probe of
systematics. In fact, as discussed by
\citet{treubrammerdiego16}, the uncertainties reported by modelers do not
include all sources of systematic errors. For example, systematic
uncertainties arising from unmodeled millilensing, residual
mass-sheet degeneracy, and multiplane lensing are very difficult to
calculate and are thus not included. The lensed-SN tests
provide estimates of the amplitude of the unknown uncertainties.
Other known sources of errors are not included either. For example, a
3\% uncertainty in the Hubble constant \citep{Riess:2011} implies a
3\% uncertainty in time delays (i.e., $\sim10$ days for a
year-long delay). Furthermore, the uncertainties are typically highly
non-Gaussian, so the 95\% confidence interval is not simply twice
as wide as the 68\% one.

\section{Conclusions}
\label{sec:summary}

With models of the \macs1149\ galaxy-cluster potential, the appearance of SN Refsdal in November 2014 as an Einstein cross became an augury of its future arrival $\sim8''$ away in a different image of its host galaxy. 
The detection of the reappearance here shows the power of modern-day predictions using models of the distribution of matter in galaxy clusters and the general theory of relativity. 
The timing and brightness of light from \snrefsdal\ in image SX is approximately in agreement with predictions, 
implying that for most models, unknown systematic uncertainties cannot be substantially larger
than random uncertainties. 
At the same time, this first detection provides some
discriminatory power: not all models fare equally well. Grillo-g, Oguri-g,
Oguri-a, and Sharon-a appear to be the ones that match the observations most closely. In
general, most models seem to predict a slightly higher magnification
ratio than observed, or shorter delays. 

From the light curves of images S1--S4 of SN Refsdal, we can already anticipate how the brightness of 
image SX will evolve.  An {\it HST} imaging program will continue to measure 
the light curve of image SX past peak brightness through 2017 (PI Kelly; GO-14199)\footnote{Indeed, as we will report in detail elsewhere, {\it HST} data obtained in January 2016 show SX brightening.}
 and constrain the time delay and magnification of image SX relative to images S1--S4 to within 1--2\%.  
These measurements will make it possible to discriminate among the model predictions with improved precision.

\acknowledgements

We express our appreciation for the efforts of Program Coordinator
Beth Periello and Contact Scientist Norbert Pirzkal of STScI.
Support for the analysis in this paper is from {\it HST} grant
GO-14041. The GLASS program is supported by GO-13459, and the
FrontierSN photometric follow-up program has funding through GO-13386.  
A.Z. is supported by Hubble Fellowship (HF2-51334.001-A) awarded by STScI, which is operated for NASA by
the Association of Universities for Research in Astronomy, Inc. under 
contract NAS 5-26555.  R.J.F. gratefully acknowledges support
from NSF grant AST-1518052 and the Alfred P.\ Sloan Foundation.  A.V.F.'s
group at UC Berkeley has received generous financial assistance from
the Christopher R. Redlich Fund, the TABASGO Foundation,
and NSF grant AST-1211916. M.N. acknowledges PRIN-INAF 2014 1.05.01.94.02.  
This supernova research at Rutgers University is supported by NSF CAREER award AST-0847157, as well as NASA/Keck JPL RSA 1508337 and 1520634, to S.W.J.

\end{document}